\documentclass[prl,twocolumn,showpacs,floatfix]{revtex4}
\usepackage{graphicx}

\begin{document}

\title{Magic Number $7 \pm 2$ in Globally Coupled Dynamical Systems?}

\author{Kunihiko Kaneko}
\email{kaneko@complex.c.u-tokyo.ac.jp}
\affiliation{Department of Basic Science, School of Arts and Sciences, University of Tokyo,
3-8-1 Komaba, Meguro, Tokyo 153-8902, Japan}

\date{\today}

\begin{abstract}

The prevalence of Milnor attractors has recently been reported in a class of
high-dimensional dynamical systems.  We study how this prevalence depends on
the number of degrees of freedom by using a globally coupled map and
show that the basin fraction of Milnor attractors increases drastically
around 5-10 degrees of freedom,  saturating for higher numbers of
degrees of freedom.  It is argued that this dominance of Milnor attractors
in the basin arises from a combinatorial explosion of the  basin boundaries.
In addition, the dominance is also found
in a system without permutation symmetry, i,e., a coupled
dynamical system of non-identical elements.  Possible relevance to the
magic number $7\pm 2$ in psychology is discussed.
\end{abstract}


\maketitle

In recent studies of dynamical systems with many
degrees of freedom, the prevalence of Milnor attractors has been recognized\cite{KK-Milnor,pulse}.
The  Milnor attractors are defined as follows:
an arbitrary small perturbation to an orbit at a Milnor attractor
can kick the orbit away from it to a different attractor, even though
a finite measure of initial conditions is attracted to the attractor by temporal
evolution\cite{Milnor,Ashwin,riddled}.  In other words, the basin of the attractor touches
the attractor itself somewhere. An orbit is often attracted to
a Milnor attractor, but can be kicked away from it by infinitesimal
perturbation.

It should be noted that Milnor attractors can exist in low dimensional
dynamical systems like a two-dimensional map as well.
When changing the parameter of a dynamical system, the basin boundary of an
attractor may move until, for a specific value of the parameter, the 
basin boundary touches the attractor.
If, for this parameter value,  the attractor has
a positive measure of initial conditions forming the basin of attraction, 
it becomes a Milnor attractor.  
It is naively expected that the
above situation occurs only for very specific parameter values (e.g., at a bifurcation point),
and that the Milnor attractors may not exist
for an open set of parameter values.

In recent studies of a class of dynamical systems, however, 
Milnor attractors are found to be sometime prevalent, occurring not only for specific
isolated parameter values,
but for continuous ranges of parameter
values.  Furthermore, the measure of the initial conditions
that belong to the basin of these Milnor attractors is a relatively large
proportion of the phase space.  Indeed, for some parameter ranges,
almost all randomly chosen initial conditions 
fall onto Milnor attractors\cite{KK-Milnor,pulse}.

Such dominance of Milnor attractors is often found in high-dimensional
dynamical systems, for example. globally coupled maps with 10
degrees of freedom or so, and within a range of parameter values where
many attractors coexist.
The question we address in the present Letter is why
can there be so many Milnor attractors in a ``high-dimensional''
dynamical system, and how many elements are
sufficient for constituting such `high' dimensionality.
With the help of numerical results
obtained from simulations of globally coupled maps,
we will show that the dominance starts to be common
at $5 \sim 10$ degrees of freedom.
We propose a possible origin for this dominance of Milnor attractors,
by noting a combinatorial explosion of
saddle points (or more generally basin boundary points).
We will also show that the prevalence of Milnor attractors
is observed even in a system without symmetry, i.e., in a coupled dynamical system
of non-identical elements.
%, by taking a coupled map whose elements have non-identical parameters.
Finally we briefly discuss the possible relevance of
our results to the magic number $7\pm 2$ discussed in psychology.

As a prototype example to study this problem we use a
globally coupled map\cite{KK-GCM}

\begin{equation}
x_{n+1}(i)=(1-\epsilon) f(x_n(i))+\frac{\epsilon}{N}\sum_j f(x_n(j)),
\end{equation}

\noindent
where $n$ is the discrete time  and $i$  the index for
its elements ($i= 1,2, \cdots ,N$ = dimension of the system).
For the elements we choose $f(x)=1-a x^2$, since the model has thoroughly been
investigated
as a prototype model for high-dimensional dynamical systems.
The coupling parameter $\epsilon$ is fixed at 0.1, since for this value
the typical behaviors of the above GCM that are relevant here
can be observed by changing only $a$.

In the present model, each attractor can be coded by the so-called
clustering condition, that is to say, by the way how the $N$ elements of the
system partition into
mutually synchronized clusters, i.e., a set of elements in which $x_n(i)=x_n(j)$
\cite{KK-GCM}. Each attractor is coded by the number of clusters $k$ and the
number of elements $N_k$ in each cluster with the
clustering condition given by $(N_1,N_2,\cdots,N_k )$\cite{rem-band}.
Indeed, for a GCM with $N=10$\cite{KK-Milnor}, that many initial conditions are
attracted to a Milnor attractor.
Here, in order to discuss the dependence of the size of the basin fraction of the Milnor
attractors on the number of  degrees of freedom, we have computed 
the ratio of initial conditions that are attracted to Milnor attractors.

\begin{figure}
%\noindent
%\hspace{-.3in}
%\includegraphics[width=58mm,angle=-90]{fig1.ps}
%\includegraphics[width=58mm]{fig1.eps}
\includegraphics[width=84mm]{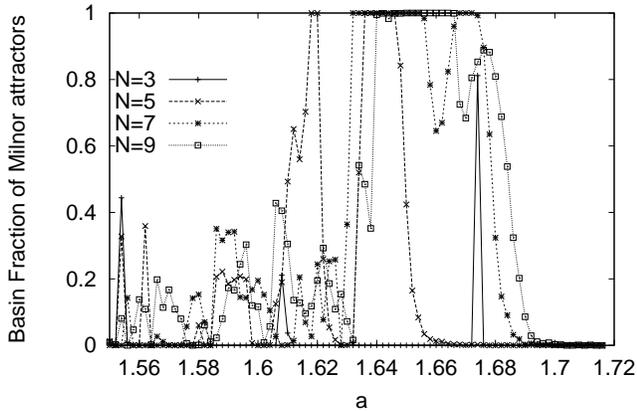}
\caption{
The basin fraction of Milnor attractors plotted as a function of the
parameter $a$, for $N=3,5,7,$ and 9.  For the present 
simulations, we take 1000 randomly chosen initial conditions,
and iterate $10^5$ steps.  Then the orbit is perturbed
as $x_n(i)+10^{-10} \sigma$, with $\sigma$ as a random number over 
[-.5,.5].  With 1000 trials of such perturbations, we checked whether
the orbit remains on the same attractor or not.  
If some of the 100 trials result in an escape from the original attractor,
it is regarded as a Milnor attractor\cite{rem00}.  
}
\end{figure}

As shown in Fig.1, the basin fraction of Milnor attractors is large around $a \approx 1.65$,
especially for $N \geq 5$.  Its parameter dependence, however,
is quite strongly dependent on $N$,
at least when $N$ is not so large.  
Hence, it is not that relevant to compare the behaviors for different $N$ 
at a given value of $a$.  Instead, we compute the average 
basin fraction of Milnor attractors
over the parameter interval $1.55<a<1.72$.  In Fig.2, this average is plotted
 as a function of the number of degrees of freedom $N$.
The increase of the average basin fraction of Milnor attractors with $N$ is clearly visible
for $N \approx (5 \sim 10)$, while it levels off for 
$N>10$\cite{Rem}.

\begin{figure}
%\noindent
%\hspace{-.3in}
%\includegraphics[width=58mm,angle=-90]{fig2.ps}
%\includegraphics[width=58mm]{fig2.eps}
\includegraphics[width=84mm]{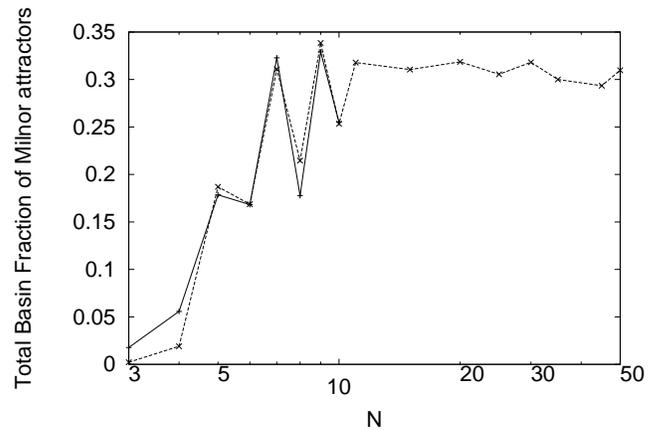}
\caption{
The average fraction of the basin ratio of Milnor attractors.
After the basin fraction of Milnor attractor is computed 
as in Fig.1, the average of the ratios for parameter values 
$a=1.550,1.552,1.554, \cdots 1.72$ is taken.
This average fraction is plotted as a function of $N$.  
(The dotted line depicts the case where $a$ is incremented by
0.01 instead of 0.002; For $N>10$, the simulation is carried out
only with this choice.)
}
\end{figure}

Now we discuss how the dominance of Milnor attractors appears.
%might lie in combinatorial explosion of attractors.  
In a system with identical elements, due to the symmetry, there are at least
\begin{math} 
M(N_1,\cdots ,N_k)=\frac{N!}{\prod_{i=1}^{k} N_i !} 
\prod_{\mathrm{over sets of} N_i=N_j}\frac{1}{m_{\ell}!} 
\end{math}
attractors for each clustering condition, where $m_{\ell}$ is the
number of clusters with the same value $N_j$\cite{KK-Milnor}.
Then, combinatorial increase in the number of attractors can be expected
when many of the clustering conditions are allowed as attractors.
%For example, the permutation of $N$ elements leads to $(N-1)!$ possibilities %and one might expect the number of  attractors to be of this order.  
%On the other hand, the phase space volume in a coupled
%system expands only exponentially with $N$.  Typically the combinatorial explosion
%outruns the exponential increase around $N \approx (5 \sim 10)$.
%(For example compare $2^N$ and $(N-1)!$.  The latter surpasses the former
%at $N=6$\cite{comb}.)  Hence, the attractors crowd the phase space\cite{Wiesenfeld},
%and the stability of each attractor may be lost.
However, the increase of the number of attractors cannot explain
the increase of the basin for Milnor attractors.
In Fig.3. we have plotted the number of attractors and the basin fraction
of Milnor attractors for $N=10$.  As can clearly be seen, 
the dominance of the Milnor attractors is not necessarily
observed when the number of attractors is high. Rather, 
the fraction of Milnor attractors gets large even when many
attractors start to disappear with the increase of $a$.%\cite{change}.
%Since the basin volume of each of the attractors is far from being equal,
%the explosion in the number of attractors does not necessarily mean that
%the basin volume for each and every attractor should be very small.
%Indeed, according to our numerical results, for the parameter region where
%the Milnor attractors dominate, the number of Milnor attractors is not
%so high and the basin fraction of only a few Milnor attractors occupies
%almost all of phase space.

\begin{figure}
%\noindent
%\hspace{-.3in}
%\epsfig{file=fig3.ps,width=.7\textwidth,angle=-90}
%\includegraphics[width=58mm,angle=-90]{fig3.ps}
\includegraphics[width=84mm]{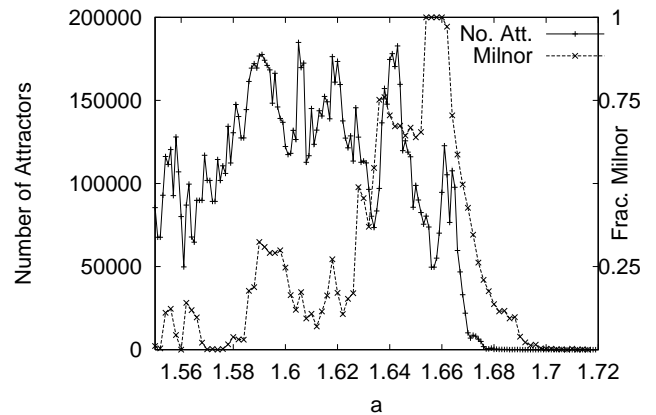}
\caption{
The number of attractors (+) estimated
from simulations over $10^5$ initial conditions.  
The estimated number of attractors is
plotted as a function of $a$. $N=10$.
All attractors that are concluded to exist by the symmetry 
argument are also counted.
The basin fraction of Milnor attractors obtained in the same way
as in fig.1 is also by dotted line with $\times$.
}
\end{figure}

%In the above sequence, the dominance of the Milnor attractors is observed
%when many attractors have disappeared.  Therefore, we can revise the
%first explanation on the dominance by replacing the combinatorial explosion
%in the number of attractors themselves by the combinatorial explosion
%in the number of basin boundaries that
%separate the  attractors.  
For the parameter region where many attractors
start to disappear, there remain basin  boundary points separating such 
(collapsed) attractors and the remaining attractors.  
To explain the prevalence of Milnor attractors, we discuss how the 
distance between an attractor and its basin boundary changes with $N$.
Consider an interval having one basin boundary for $x_1>x^*$ and $x_1<x^*$ separating two 
attractors. If an $N$-dimensional phase space 
is partitioned by (basin) boundaries at $x_i=x^*$ for $i=1,\cdots N$,
it is partitioned into $2^N$ units.  In this case,
the distance between each attractor and the basin boundary
does not change with $N$.

On the other hand, consider a boundary given by some condition for 
$(x_1,\cdots , x_N)$.  In the present system with  global (all-to-all) couplings,
many of permutational change of $x_i$ in the condition 
give also basin boundaries.  Here the condition for the basin
is typically related with some attractor (or a ruin of collapsed attractor) that has 
clustering $(N_1,\cdots ,N_k)$.  Then
there are $M(N_1,\cdots ,N_k)$ partitions by boundaries equivalent by permutations.
The number of regions partitioned by the boundaries   
increases combinatorially with $N$.  Roughly speaking, it increases
in the order of $(N-1)!$, when the boundary has
a variety of clusterings (i.e., large $M$).
Recalling that the distance between an attractor and the basin boundary
remain at the same order for the partition of the order of $2^N$, 
the distance should decrease if $(N-1)!$ is larger than $2^N$.  Since the
former increases remarkably faster for $N>5$ than the latter,
the distance should decrease drastically for $N>5$.
Since this argument is applied for any basin boundary condition with
a complex clustering having combinatorially large $M(N_1,\cdots ,N_k)$,
the distance between an attractor and such basin boundary 
may often go drastically small.  For a certain parameter regime
( $1.64< a <1.68$ in the present case),
basin boundaries with such partitions are dominant, and the above argument is applied
to most attractors.
Although this explanation may be rather
rough, it gives a hint to why Milnor attractors
are so dominant when $N$ is larger than $5 \sim 10$.
%We surmise that this is the reason why
%Milnor attractors are dominant in our model 
%at $(1.64 \sim 1.67)$ when $N \stackrel{>}{\approx}(5 \sim 10)$.

Since the above discussion is based mainly on simple combinatorial
arguments, what we need is instability in orbits leading to many attractors,
and global (all-to-all) coupling to allow for permutations of elements.
Then, the dominance of Milnor attractors
for $N \stackrel{>}{\approx}(5 \sim 10)$ may be rather common in
globally coupled dynamical systems.  To check this possibility, we have also made
numerical simulations for Josephson junction series arrays that are globally coupled
through a resistive shunting load and driven by an rf bias current\cite{JJ}, given by
\begin{math}
\ddot{\phi _j}+g\dot{\phi _j} +sin\phi _j+g\sigma/N \sum_{m=1}^N \dot{\phi_m}=
i_{dc}+i_{rf}sin(\Omega_rf t).
\end{math}
By using the parameter values adopted in \cite{JJ}, we have computed
the basin volumes for Milnor attractors, at the partially ordered phase
where a variety of attractors with many clusters coexists.  Again,
the basin volumes are close to 0 for $N \leq 4$, and increase at $5<N<10$, 
%to reach a large portion for $N>10$.
It is also interesting that
pulse-coupled oscillators with global coupling also show
the prevalence of Milnor attractors for $N \geq 5$
\cite{pulse}.

So far reports on the prevalence of Milnor attractors have been limited
to a system with symmetry.  For example, in the GCM (1),
the permutation symmetry arising from employing identical elements leads to a
combinatorial explosion in the number of attractors as mentioned.
Then, one may wonder whether the prevalence of Milnor attractors 
is possible only for such highly symmetric systems,
%especially since it has been demonstrated that some Milnor attractors  
especially because some Milnor attractors are known to 
disappear when introducing  tiny asymmetries\cite{Ott}.
%Hence it is important to check whether the prevalence of Milnor attractors
%can be observed in a system without symmetry.  
We have therefore studied a GCM with 
inhomogeneous parameters\cite{hetero}, given by
\begin{math}
x_{n+1}(i)=(1-\epsilon) f_i(x_n(i))+\frac{\epsilon}{N}\sum_j f_j(x_n(j)).
\end{math}
%\noindent
with $f_i(x)=1-a_i x^2$, and
$a_i=a_0+a_w \frac{i-1}{N-1}$.

In Fig.4, we have plotted the basin fraction for Milnor attractors 
with the change of the parameter $a_0$ while fixing $a_w=0.1$, for
$N=3,4,\cdots,10$.  
Although the fraction is smaller than in the homogeneous case,
Milnor attractors are again observed and found to
dominate the basin volume for some parameter region. 
As in the symmetric case, the basin fraction of Milnor attractors 
increases around $N \approx (5 \sim 10)$.

\begin{figure}
%\noindent
%\hspace{-.3in}
%\epsfig{file=fig4.ps,width=.7\textwidth,angle=-90}
%\includegraphics[width=58mm,angle=-90]{fig4.ps}
%\includegraphics[width=58mm]{fig4.eps}
\includegraphics[width=84mm]{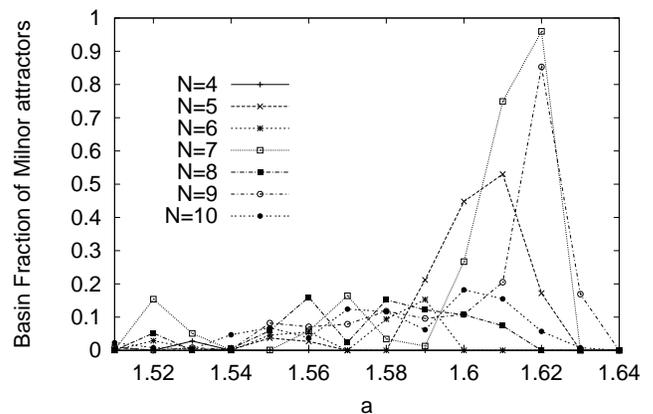}
\caption{
The basin fraction of Milnor attractors for a GCM with
inhomogeneous parameters.
Here $a(i)=a_0+.1 \times (i-1)/(N-1)$. Plotted as a function of the
parameter $a_0$, for $N=4-10$. (For $N=3$, the fraction is almost zero
for all $a_0$.)  
The fraction is computed in the same way as in Fig.1, except that 100 trials of 
perturbations were used
instead of 1000.  Since the clustering condition cannot be used in this case,
%we adopted the following procedure for
%distinguishing the attractors.
%To check whether the orbit is on the same attractor, we computed the temporal 
we checked whether the orbit is on the same attractor, by computing the temporal 
average of $x_n(i)$ over $5\times 10^6$ steps,
before and after each perturbation.  If the average agrees to within 
a precision of $10^{-3}$, the orbit is regarded to be on the same attractor.
(We have investigated other average times in the range from $10^4$ to
$10^5$ with  precisions ranging from $10^{-4}$ to $10^{-2}$.  
Even though the basin fractions
may change a little, the existence of the 
Milnor attractors and their dominance around $a_0=1.61$ for $N=7,9$ 
is invariant.)}
%Inset: the accumulated fraction of Milnor attractors for $1.52<a_0<1/71$ .
\end{figure}

Note that even though complete synchronization between two elements is lost,
clusterings as with regards to the phase relationships can exist\cite{Pikovsky}.
For example, there are two groups when considering the oscillations of phases
as large-small-large... and small-large-small..., that
are preserved in time for many attractors. 
%Furthermore, finer preserved phase relationships can also exist.  
Similarly, it is natural to expect an explosion in the number of the basin boundaries
for some parameter regime.  Accordingly the argument
on the dominance of Milnor attractors for a homogeneous GCM 
can be applied here to some degree as well.

The term magic number $7 \pm 2$ was originally coined in psychology
\cite{Miller}, when it was found that 
the number of chunks (items) that is memorized in short term memory
is limited to $7 \pm 2$.  (See \cite{TsudaNicolis} for a pioneering study
to relate this number with chaos.)
%the number 7 crucially depends on the choice of specific parameters in the model so far. 
%On the other hand, 
%With combinatorial arguments, it can be shown that
%our `magic number $5 \sim 10$' in dynamical systems
%does not strongly depend on the choice of parameters, and it may be interesting to 
%discuss a possible connection with the original magic number $7 \pm 2$ in psychology.
To memorize $k$ chunks of information including their order (e.g.,
a phone number of $k$ digits) within a dynamical system, 
let us assign each memorized state to an
attractor of a $k$-dimensional dynamical system, as is generally
adopted in neural network studies.
In this $k$ dimensional phase space, a combinatorial variety of attractors
has to be presumed in order to assure a sufficient variety of memories.  Depending on the
initial condition (given by inputs), an orbit has to be separated to
different attractors.  Then, a combinatorial explosion
of basin boundaries is generally expected with the increase of $k$, if
the neural dynamics in concern are globally coupled (as often 
adopted in neural networks).  Then, following
the argument in the present paper, Milnor attractors may be dominant 
for $k > (5 \sim 10)$.  (Recall that the number does not strongly depend on the choice of
models, since it is given by the combinatorial argument.)
Since the state represented by a Milnor attractor is
kicked out by tiny perturbations, robust
memory may not be possible for information that contains more than 
$7 \pm 2$\cite{rem3} chunks.
Although this explanation is a rough sketch, it
can possibly be applied to
other systems that adopt attractors as memory\cite{rem5}.
 
In dynamical systems, it is well known that the dimensional cutoff $\geq 3$
plays an important role for the existence of chaos.
It is interesting then to investigate whether there are certain higher 
dimensions that similarly form dimensional boundaries 
beyond which the behavior
of a dynamical system changes qualitatively.  The present study may
shed new light on this possibility.  Also, it is interesting to note that in 
Hamiltonian dynamical systems, agreement with thermodynamic behavior is often
observed only for degrees of freedom higher than $5 \sim 10$\cite{Sasa}.  
Considering the combinatorial complexity woven by  all the possible Arnold webs 
(that hence may be termed ``Arnold spaghetti''), the entire phase space volume 
that expands only exponentially with the number of degrees of freedom 
may be covered by webs, resulting in
uniformly chaotic behavior.  If this argument holds, the
degrees of freedom required for thermodynamic behavior can also be
discussed along the line of the present Letter.

\begin{acknowledgments}
I would like to thank F. Willeboordse for critical reading of the manuscript.
%M Timme for sending me \cite{pulse} prior to publication,
%and Y. Takahashi for illuminating discussions.
This work is partially supported by a Grant-in-Aid for Scientific
Research from the Ministry of Education, Science, and Culture
of Japan.
\end{acknowledgments}


\begin{thebibliography}{9}
%\addcontentsline{toc}{section}{References}

%\begin{thebibliography}{999}

\bibitem{KK-Milnor}
K. Kaneko, Phys. Rev. Lett., 78 (1997) 2736;
Physica D, 124 (1998) 322.

\bibitem{pulse}
M. Timme, F. Wolf,and  T. Geisel., `Prevalence of unstable attractors in
networks of pulse-coupled oscillators' preprint.

\bibitem{Milnor}
J. Milnor, Comm. Math. Phys. 99 (1985) 177;  102 (1985) 517.

\bibitem{Ashwin}
P. Ashwin, J. Buescu, and I. Stuart,
Phys. Lett. A 193 (1994) 126;
Nonlinearity 9 (1996) 703.

\bibitem{riddled}
J.C. Sommerer and E. Ott., Nature 365 (1993) 138;
E. Ott et al., Phys. Rev. Lett. 71 (1993) 4134;
Y-C. Lai and C. Grebogi, Phys. Rev. E. 53 (1996) 1371.

\bibitem{KK-GCM}
K. Kaneko, Phys. Rev. Lett. 63 (1989) 219; Physica 41D (1990) 38.

\bibitem{rem-band}
For some cases it is also necessary to establish whether an attractor
belongs to a period-2 band in the logistic map, since
for some parameter region, the period-2 band motion remains even though the
motion itself is chaotic\cite{KK-Milnor}.
%For example, a clustering condition for complete de-synchronization
%(i.e., (1,1,1,...,1) ) can have more than one attractor when the
%motion of each element obeys the period-2 band.  Elements split into two groups
%i.e. large-small-large-small and small-large-small-... oscillations.

\bibitem{rem00}
In the estimate here, `attractors' with an escape rate of
more than 50\% are not counted as Milnor attractors in order to avoid possible
inclusion of transient states. Even if they are included, however, there is 
only a small increase in the basin fraction.  Note that in spite of the `severe' 
criterion adopted here, Milnor attractors are dominant. See also [1].
%the dominance of Milnor attractors is shown.

\bibitem{Rem}
In this class of models, we have found that the fraction of Milnor attractors
is larger for odd $N$ than for even $N$.  Note that  two clusters with
equal cluster numbers and anti-phase oscillations generally have less chaotic instability.  
A globally coupled map with an even number of elements
allows for equal partition into two clusters.  This gives a plausible
explanation for the smaller instability for even $N$ system.

%\bibitem{Wiesenfeld}
%The idea of this type of attractor crowding was first proposed by 
%P. Hadley and K. Wiesenfeld (Phys. Rev. Lett. 62 (1989) 1335).  
%However, it was later shown that states with different
%phase orderings  (with $(N-1)!$ variety) are not separate attractors
%(K. Kaneko, Physica  55D (1992) 368; S. Watanabe and S. Strogatz, Phys. Rev. Lett. 70 (1993) 2391).
%these proved by S. Watanabe and S. Strogatz, Phys. Rev. Lett. 70 (1993) 2391 that states with different
%phase orderings  (with $(N-1)!$ variety)  are not separate attractors but  reside on the same torus, 
%as also pointed out by K. Kaneko, Physica  55D (1992) 368.

%\bibitem{comb}
%Of course, this estimate depends on the way in which the  phase space volume $\alpha ^N$ is increased.  
%For $\alpha =1.5$ the crossover occurs at about $N=4$, for $\alpha = 3$ at $N \approx (6 \sim 7)$, for $\alpha =4$,
%$N \approx 11$.  At any rate, the crossover size does not show a sharp increase with $\alpha$.

%\bibitem{change}
%From numerical simulations, the behavior of the GCM (1) changes as follows with the increase of the 
%parameter $a$; {\sl a few attractors with small numbers of clusters;} $\rightarrow$ 
%{\sl increase of the number of attractors with stable and Milnor attractors coexisting;} $\rightarrow$ 
%{\sl decrease of the number of attractors with some remaining Milnor attractors with large basin fractions;}
%$\rightarrow$ {\sl only a single or a few stable  attractors with complete de-synchronization.}

\bibitem{JJ}
D.Dominguez and H.A. Cerdeira, Phys. Rev. Lett. 71 (1993) 3359

\bibitem{Ott}
S.C. Venkataramani, B.R. Hunt, and E. Ott,
Phys. Rev. E 54 (1996) 1346.

\bibitem{hetero}
K.Kaneko, Physica 75 D (1994) 55.

\bibitem{Pikovsky}
%In simple synchronization, the synchronization of the phase only
Synchronization of the phase only
is known as phase synchronization;
M.G. Rosenblum, A.S. Pikovsky, and K. Kurths,
Phys. Rev. Lett 76 (1996) 1804 ( see also
K. Kaneko, Physica 37D (1989)60).  
Clusterings only as to the phases of oscillations are its natural
extension.

\bibitem{Miller}
G.A. Miller, {\sl The psychology of communication}, 1975, Basic Books, N.Y.

\bibitem{TsudaNicolis}
J.S.Nicolis and I.Tsuda,
%"Chaotic dynamics of information processing: The magic number seven plus-minus two revisited, 
Bull. Math. Biol.  47(1985)343.


\bibitem{rem3}
This does not necessarily imply that Milnor attractors are irrelevant to
cognitive processes.  Milnor attractors, for example, may be useful for
 chaotic search processes\cite{KK-Milnor}.  Searches with chaos itinerating 
over attractor ruins has been discussed in\cite{Freeman,Tsuda} with an
experimental support on the olfactory bulb\cite{Freeman}.

\bibitem{Freeman}
W. Freeman and C. A. Skarda,  Brain Res. Rev. 10 (1985) 147; Physica D 75
(1994)151.

\bibitem{Tsuda}
I. Tsuda, Neural Networks 5(1992)313.

\bibitem{rem5}
Of course, the present argument should mainly be applied to systems with 
all-to-all couplings, or to highly connected network systems.
If the connections are hierarchically ordered, the number of memory items can be
increased.  The often adopted module structure is relevant for this purpose.

\bibitem{Sasa}
S. Sasa and T.S. Komatsu, Phys. Rev. Lett. 82 (1999) 912:
N. Nakagawa and K. Kaneko, Phys. Rev. E 64(2001) 055205(R)-209:

%and unpublished data.

\end{thebibliography}
\end{document}